\documentclass[12pt]{iopart}

\usepackage{iopams}
\usepackage{graphicx}

\begin{document}

\title[Transient eddy current flow metering]{Transient 
eddy current flow metering}

\author{J Forbriger and F Stefani}

\address{Helmholtz-Zentrum Dresden - Rossendorf, PO Box 510119,
D-01314 Dresden, Germany}
\ead{F.Stefani@hzdr.de}
\begin{abstract}

Measuring local velocities or entire 
flow rates in liquid metals or semiconductor melts 
is a notorious problem in many industrial applications,
including metal casting and silicon crystal growth.
We present a new variant of an old
technique which relies on the 
continuous tracking of a flow-advected 
transient eddy current that is induced by a pulsed
external magnetic field.
This calibration-free method 
is validated by applying it
to the velocity of a spinning disk made of 
aluminum. 
First tests at a rig with a flow of liquid 
GaInSn are also presented.

\end{abstract}

\vspace{2pc}
\submitto{\MST}
\maketitle

\section{Introduction}

Compared to transparent fluids, liquid metals
and semiconductor melts pose serious problems
for the reliable determination of local velocities 
or integral flow characteristics \cite{ECKERT}. Even the widely
used ultrasonic Doppler velocimetry (UDV) faces 
difficulties when it comes to its application 
in very hot and/or 
chemically aggressive fluids, such as liquid steel 
or silicon. 
Fortunately, the high electrical conductivity
that is responsible for the opaqueness 
of those fluids allows the  
utilization of magnetic inductive methods.

These methods bear on 
applying magnetic fields to the flowing fluid and
measuring appropriate features, e.g. amplitudes, phases, 
or forces, of the flow induced magnetic fields. 
In the contactless
inductive flow tomography (CIFT), entire
three-dimensional flow fields are reconstructed 
from induced field amplitudes that are measured 
at many position around the fluid, to which one or
a few external magnetic fields are applied 
\cite{MST1,CIFT}. A recently developed
flow rate sensor relies on the determination
of magnetic phase shifts due to the flow 
\cite{PRIEDE,BUCHENAU}. In the Lorentz force 
velocimetry (LFV) \cite{THESS}
one measures the force acting on a permanent magnet
close to the flow, which results as a direct consequence
of Newton's third law applied to the braking
force acting by the magnet on the flow.
With this technique, it is now possible to measure
velocities of fluids with remarkably low conductivities,
such as salt water \cite{HALBEDEL}.

A common drawback of all these methods is that
the measured signal is not only
dependent on the sought flow velocity, but also
on the conductivity of the fluid. 
Usually, 
the signals are proportional to the so-called magnetic
Reynolds number $Rm=\mu_0 \sigma V L$, where
$\mu_0$ is the magnetic permeability constant, $\sigma$ the
conductivity of the liquid, and $V$ and $L$ denote 
typical velocity and length scales of the relevant 
fluid volume.
In most cases, the signal depends also on geometric
factors, so that the measuring system has to be calibrated
anyway. Further to this, the use of permanent magnets 
(for LVF) or of magnetic yoke materials 
(for the phase-shift method)
set limitations to the ambient temperature 
at the position of the respective sensors.

The goal of the present paper is both to circumvent 
the necessity to calibrate the measurement system, 
and to mitigate the temperature limitation 
problem. The measurement system to be presented 
belongs to a wider class of time-of flight methods that 
utilize the existence of some traceable pattern 
in the fluid which is being advected by the flow. By 
measuring the time of flight of the advected pattern 
between two positions along the flow one can infer 
the flow velocity. 

In one realization of this principle, the role of the 
pattern is played by some (unspecified) turbulence elements 
moving close to the wall of the fluid. If the flow direction 
is known, one can apply external magnetic fields at two 
positions and infer the flow speed from correlations 
of the magnetic signals induced by the turbulence elements
that are passing by \cite{JULIUS}. A similar correlation 
technique is presently under investigation 
in the time-of-flight LVF \cite{DUBOVIKOVA}.

Another realization had already been developed in the 
1960's by Zheigur and Sermons \cite{ZHEIGUR}. 
In their `pulse method', the 
role of the traceable pattern is played
by an eddy current system that is imprinted into 
the conducting medium
by switching on or off the current in one 
excitation coil, 
and by registering the flow advected pattern by 
another coil situated downstream the flow. The 
instant at which the current in this detection
coil crosses zero 
indicates the passing-by of the center of the 
eddy current system. The flow velocity is then
determined by dividing 
the distance between the two coils 
by the measured time interval.

In \cite{ZHEIGUR}, the method had been shown to allow
to measure velocities of an aluminum bar
from 20 m/s down to 1 m/s, but not significantly lower. 
One reason for this limitation was the exclusive 
focus on identifying the instant of vanishing current 
in the detection coil by means of an oscillograph. 
For small velocities (or, more exactly, for small $Rm$), 
when the decay of the eddy current
is much faster then its advection, this instant 
smears out strongly so 
that its determination becomes less and less reliable.

In the present paper, we will apply a modern data 
acquisition system and appropriate analysis methods
in order to infer the velocity from the complete data 
sets of the induced current in three detection coils, 
rather than only one particular instant in one 
single coil. 
With view on measuring the entire
transient signal after switching, we dub
this method as `transient eddy current flow metering' 
(TEC-FM).
After explaining the measuring  principle and technical 
realization, we will apply it to the determination  of the velocity
of a rotating aluminum disk, and to the flow of GaInSn in a 
circular tube. The paper closes with some outlook for 
future improvements and for possible industrial application.

\section{Measuring system and principle}

\begin{figure}
  \centering
  \includegraphics[width=12cm]{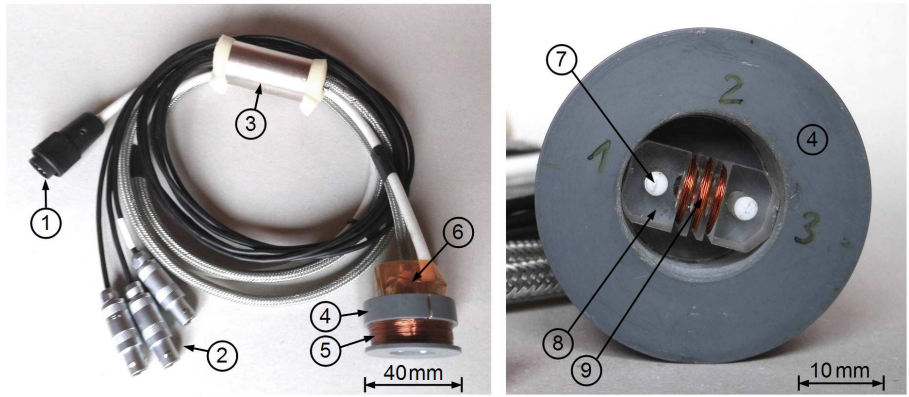}
  \caption{TEC-FM sensor with cables (left), and detailed view on 
  the coil system (right). The plugin 
  connector (1) connects the power supply to the excitation coil (5)
  consisting of 300 windings of copper wire with 0.5 mm diameter on 
  the base body (4).
  The voltages induced in the three detection coils (9) 
  are distributed in the adapter (3) to three connectors (2).
  The detection coils consist of 35 windings with diameter 0.1 mm 
  that are wound on the winding former (8) that is fixed by 
  screws (7) to the base body (4). The clamp (6) serves as a 
  cable relief.}
  \label{fig:coil}
\end{figure}

The heart of our measuring system consists of 
three small detection coils embedded into one 
larger excitation coil
(see figure \ref{fig:coil}). The excitation coil
is installed parallel to the boundary of the fluid, 
so that the three detections coils are arranged in one row
in direction of the flow.

\begin{figure}
  \centering
  \includegraphics[width=12cm]{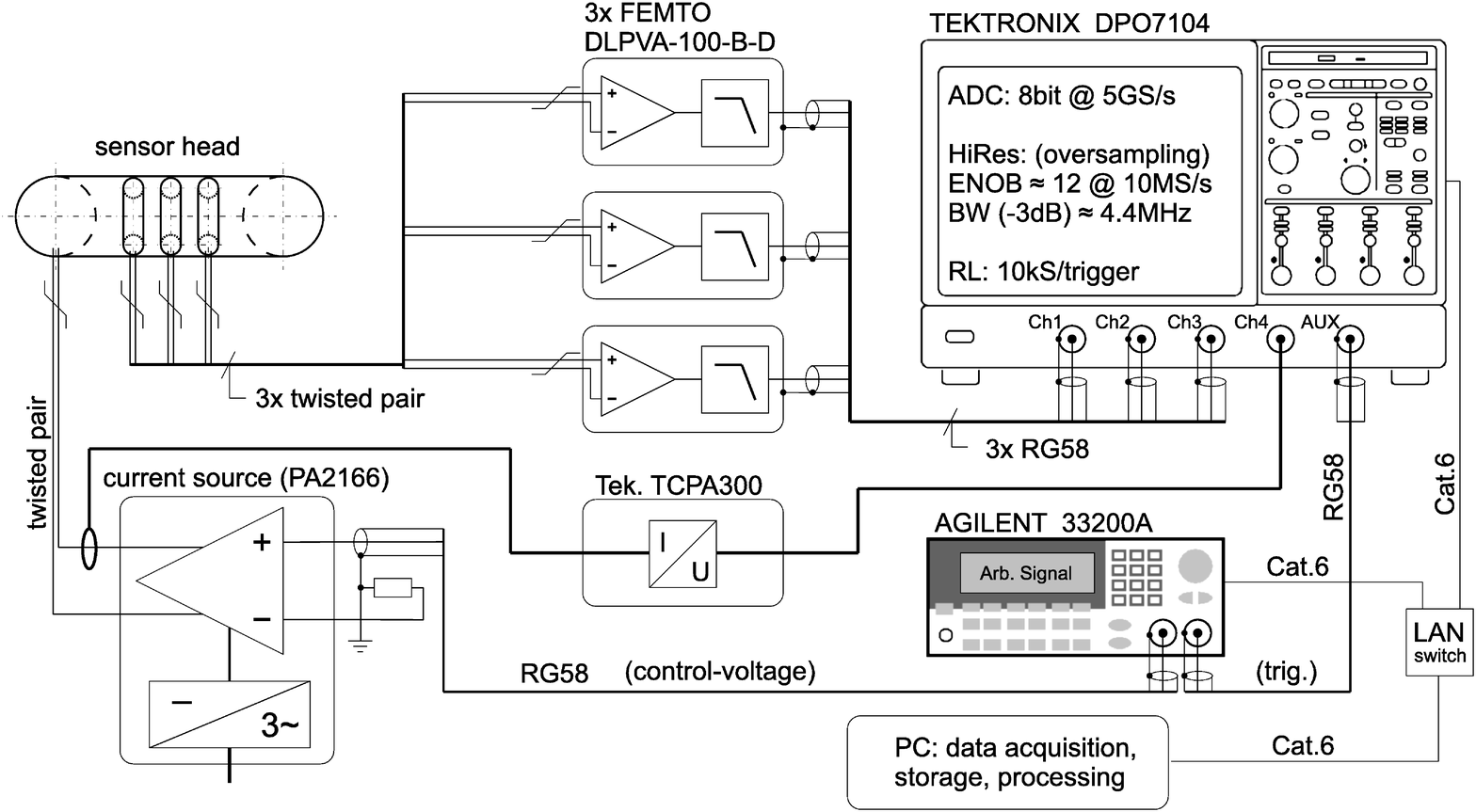}
  \caption{Block diagram of the measurement system.}
  \label{fig:block}
\end{figure}

The block wiring diagram is shown in figure \ref{fig:block}.
The signal for the excitation coil is produced 
by an AGILENT 33200A waveform generator and then  
amplified by a Rohrer PA2166 
precision power amplifier. The induced signals measured by
the three detection coils are amplified by
FEMTO DLPVA-100-B-D amplifiers and then
analyzed with a TECTRONIX DP07104 oscillograph.

\begin{figure}
  \centering
  \includegraphics[width=12cm]{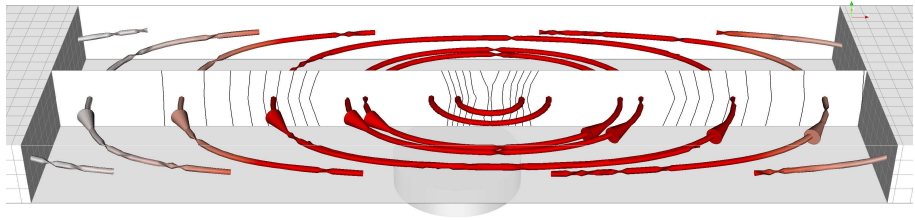}
  \caption{Simulated eddy current system induced in a right-moving
  solid body shortly after switching off the current in the 
  emitter coil. The thick red arrows in the upper panel 
  indicate the current lines, 
  the thin gray lines are iso-contours of the current density.   }
  \label{fig:eddy}
\end{figure}

In the following, we will focus on the case that 
the current in the excitation coil is on for $t<0$, 
and is being switched off at $t=0$ (note that 
the contrary case that the current is switched on 
at $t=0$ is not completely equivalent: the subtle 
differences will be discussed in the outlook). 
With our excitation technique we achieve a smooth and
non-oscillatory decay with a typical fall time of approximately
100 $\mu$s.
By this pulse, a ring-like eddy current system is induced 
in the nearby moving conductor. For a conducting half-space 
at rest, the evolution of this eddy current system  
can be given in a quasi-analytic form
\cite{EDDYBUCH}, which was also experimentally confirmed 
in a liquid metal experiment \cite{FORBRIGER}. 
Here, we illustrate the ring-like eddy current system 
for the case of a solid electrical conductor moving to the 
right. Figure \ref{fig:eddy} shows the result of 
corresponding simulation with ANSYS multiphysics. The 
excitation coil is indicated below the body.

\begin{figure}
  \centering
  \includegraphics[width=10cm]{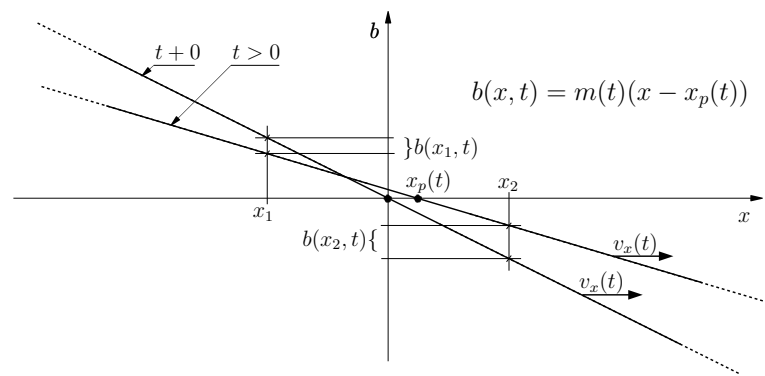}
  \caption{Schematic sketch of the evolution of the $b_x$ component
  of the magnetic field due to the eddy current.}
  \label{fig:shift}
\end{figure}

This eddy current, in turn, produces a magnetic field $\bf b$
outside the fluid, whose $b_x$-component (in flow direction) 
is zero exactly below the center of the (advected) 
eddy current ring. Close to this point, 
$b_x$ can be considered linear in $x$. 
If the detection coils are situated within this
linear region, their signals can be exploited to 
identify the point of vanishing $b_x$ which represents 
the (moving) pole of the eddy current ring. The use of three coils 
enables to verify the linearity of the function $b_x(x)$. 
Yet, for the sake of simplicity, the following
theoretical considerations will be restricted to the  
simpler two-coil system (figure \ref{fig:shift}).

Shortly after switching off the current in the excitation coil,
the $b_x$ component along the flow direction $x$ 
can be parametrized in the following way
(from here on, we replace $b_x$ by $b$):
\begin{eqnarray}
\label{eq:bxt}
b(x,t)=m(t)(x-x_p(t))
\end{eqnarray}
with the decay function
\begin{eqnarray}
m(t)=m_0 \exp{(-t/\tau)}
\end{eqnarray}
where $\tau$ is the typical decay time of the eddy
current system in the liquid. $x_p(t)$ represents the 
time-dependent pole position of the flow-advected 
eddy current system
which can, in turn, be interpreted as the time integral
over the velocity of the fluid:
\begin{eqnarray}
x_p(t)=\int_0^t v_x(t) dt  \;.
\end{eqnarray}
Hence, the velocity of the pole of the eddy-current 
related magnetic field can be determined as 
\begin{eqnarray}
v_p(t):=\frac{d x_p(t)}{d t} \; .
\end{eqnarray}
The pole position $x_p$ at a given instant 
can be determined from the values $b(x_1,t)$ and $b(x_2,t)$
measured at the two detection coil positions $x_1$ and
$x_2$, respectively:
\begin{eqnarray}
\label{eq:xp}
x_p(t)=x_1-\frac{b(x_1,t)(x_2-x_1)}{b(x_2,t)-b(x_1,t)} \; .
\end{eqnarray}
This system of equations would be 
sufficient if
the magnetic fields were indeed measured 
by Hall or Fluxgate sensors. However, since 
we are using pick-up coils
(also with view on later
high-temperature applications),
our measured signal is actually the voltage in the detection 
coil which is proportional to the time derivative of 
the magnetic fields rather than the magnetic field itself.

The equation for this time-derivative of the magnetic field,
\begin{eqnarray}
\label{eq:dotb}
\dot{b}(x,t)=\dot{m}(t)(x-x_p(t))-m(t) \dot{x}_p(t)
\end{eqnarray}
can be obtained from equation (\ref{eq:bxt}). 
The corresponding pole $x_{pp}$ of this time-derivative is 
derived from setting the r.h.s.
of  equation (\ref{eq:dotb}) to zero, and 
utilizing the relations $m(t)/\dot{m}(t)=\tau$, and
$\dot{x}_p=v_x(t)$:
\begin{eqnarray}
x_{pp}(t)=x_p(t)+\tau v_x(t) \;.
\end{eqnarray}
The velocity of this pole of the magnetic field 
derivative results than as
\begin{eqnarray}
v_{pp}:=\dot{x}_{pp}(t)=v_p(t)+\tau \dot{v}_p(t).
\end{eqnarray}
We see that the velocity of the pole of the
derivative $v_{pp}$ is identical to the velocity $v_{p}$ of the
pole of the field itself as long as the 
latter is time-independent. In our method, we will
trace indeed the pole of the time-derivative, which is
for every instant given by
\begin{eqnarray}
x_{pp}(t)=x_1-\frac{\dot{b}(x_1,t)(x_2-x_1)}{\dot{b}(x_2,t)-\dot{b}(x_1,t)} ,
\end{eqnarray}
quite in analogy with equation (\ref{eq:xp}).

\section{Rotating aluminum disks}

\begin{figure}
  \centering
  \includegraphics[width=8cm]{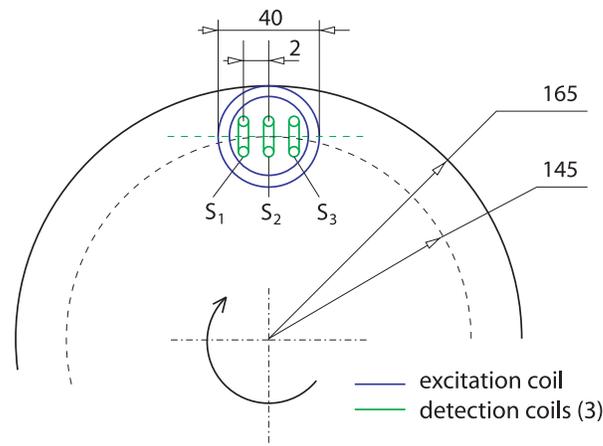}
  \caption{Velocity measurement for a rotating aluminum disk 
  (not to scale).
  }
  \label{fig:disk}
\end{figure}

In this section we will validate the TEC-FM method at 
a simple model for which the velocity is well known. For this 
purpose we chose a rotating aluminum disk of radius 165 mm and
thickness 10 mm, and apply our sensor 
at a radius of 145 mm (see figure \ref{fig:disk}).

\begin{figure}
  \centering
  \includegraphics[width=10cm]{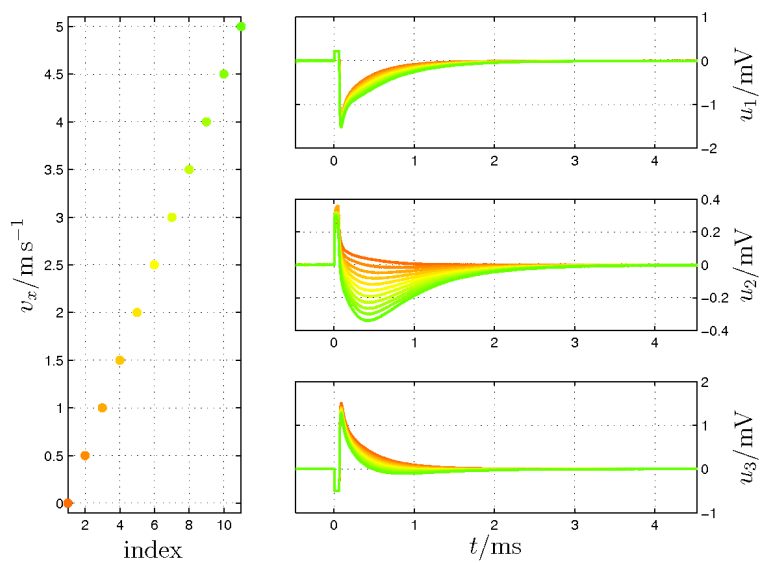}
  \caption{Time evolution of the voltages (raw signals) measured 
  in the three
  receiver coils (right) for 11 
  pre-adjusted disk velocities $v_x$ (left).
  }
  \label{fig:alu-signal}
\end{figure}

The velocity at this radius can be pre-adjusted by the disk 
rotation rate. Figure \ref{fig:alu-signal} shows  
the time evolution of the signals at the three receiver 
coils, for 11 chosen velocities between 0 m/s and 5 m/s. 

\begin{figure}
  \centering
  \includegraphics[width=9cm]{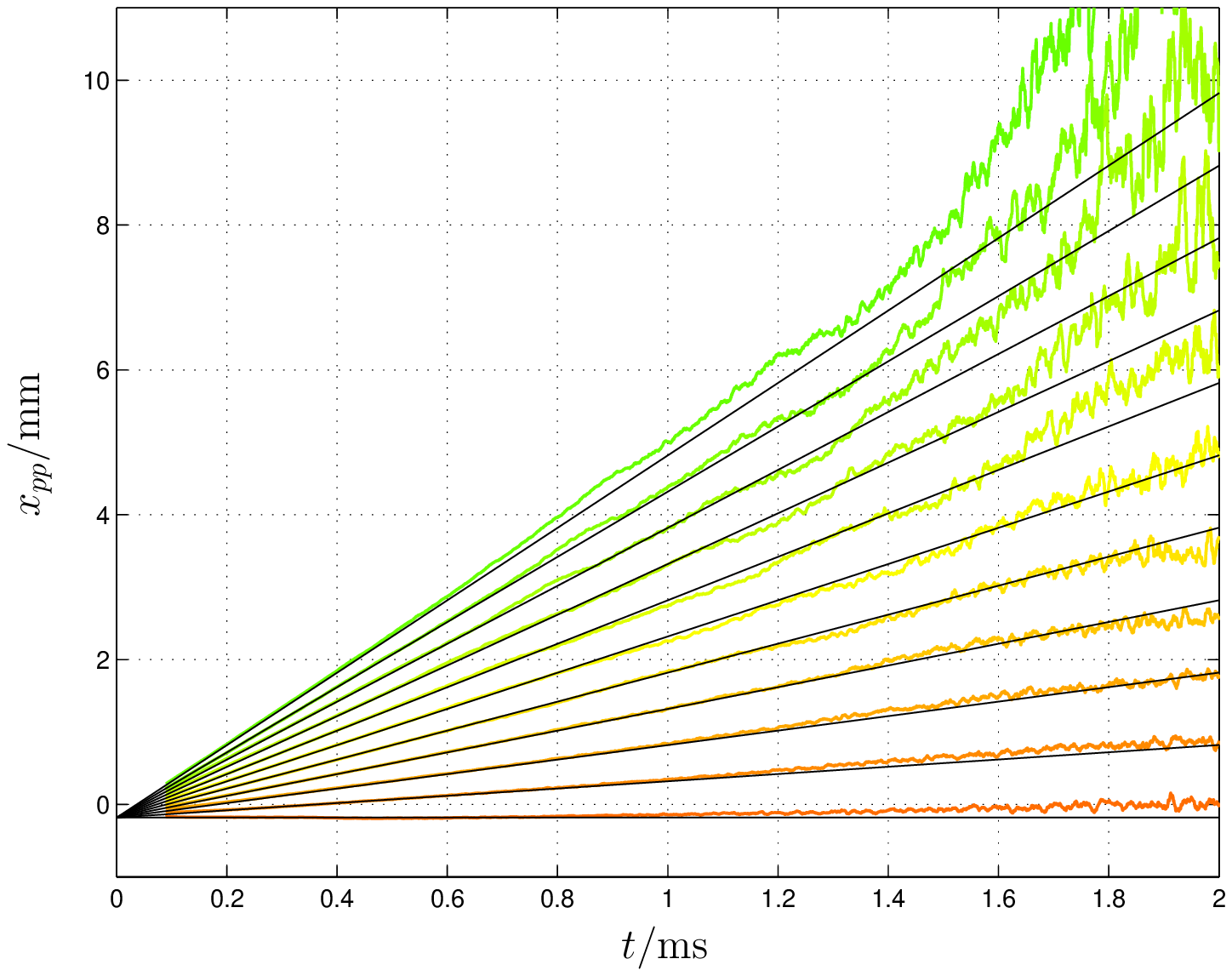}
  \caption{Time evolution of the reconstructed pole 
  position $x_{pp}$ of the eddy current for the 11 
  pre-adjusted disk velocities $v_x$.}
  \label{fig:alu-velo}
\end{figure}

From these data we estimate the position of the pole 
(figure \ref{fig:alu-velo}) for the first 2 ms after switching 
off the current. The black lines, drawn for comparison,  
would correspond to a perfect movement of the
pole of the eddy current with the rotating disk.

\begin{figure}
  \centering
  \includegraphics[width=8cm]{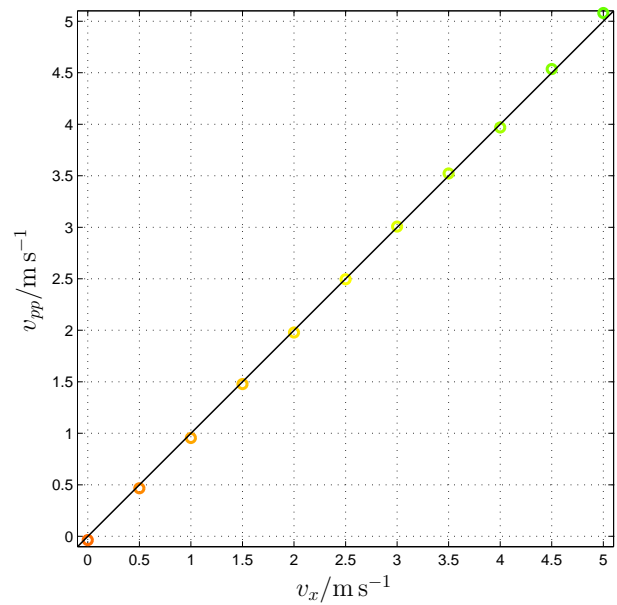}
  \caption{Inferred velocities $v_{pp}$ of the aluminum disk 
  for the 11 pre-adjusted velocities $v_x$.
  }
  \label{fig:alu-end}
\end{figure}

We see that, at least during the first 1 ms, 
the signal is very clean and allows to determine the 
speed of the pole. This is shown then in figure 
\ref{fig:alu-end}, for which we have made an 
additional 
average over 10 pulsing events with an interval of 
10 ms between them (so that we obtain one
velocity value every 100 ms). Obviously, the 
velocity can be recovered with high accuracy.

\section{Flow of GaInSn in a pipe}

\begin{figure}
  \centering
  \includegraphics[width=10cm]{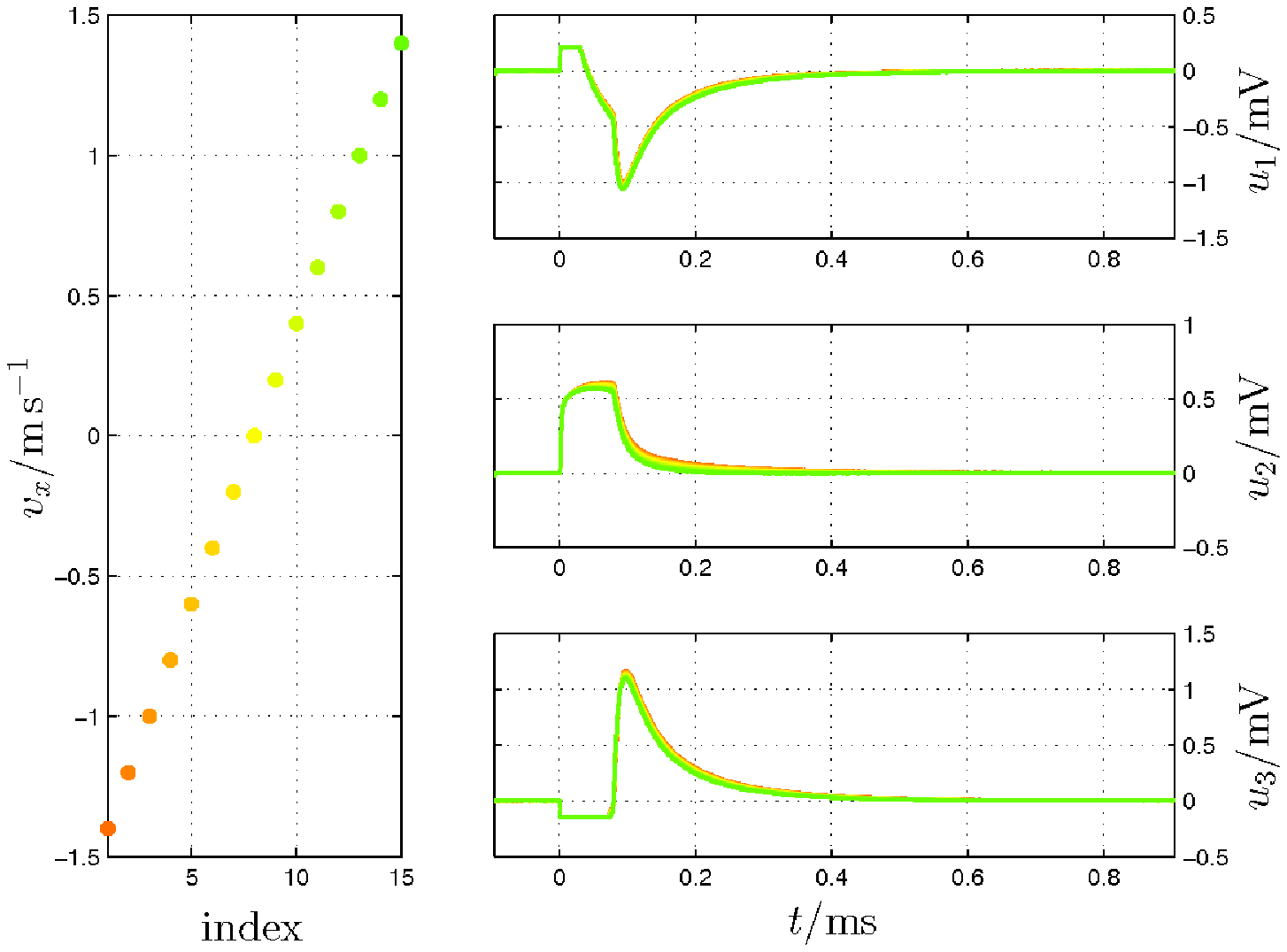}
  \caption{The same as Figure 6, but for the flow of
  GaInSn in a tube.
  }
  \label{fig:gainsn-signal}
\end{figure}

After having validated the method at a solid rotating disk, 
we apply it now to the case of a liquid metal flow in a pipe.
We consider a flow of the eutectic alloy GaInSn (with
the electrical conductivity $\sigma=3.3 \times 10^6$)
in a circular plastic pipe of inner radius 
27.3 mm (for more details of the test rig, see 
\cite{BUCHENAU}).
Figure \ref{fig:gainsn-signal} shows again the signal of 
the three coils for 15 different flow velocities 
(measured by an independent flow meter).
Note that the magnetic field decay is much faster than in the 
aluminum case with its much higher conductivity, since
the decay time is roughly $\tau=\mu_0 \sigma d^2$,
with $d$ a typical length scale of the order of the
coil distance. This lower decay time would 
allow to provide velocity values approximately 
every 10 ms.

\begin{figure}
  \centering
  \includegraphics[width=9cm]{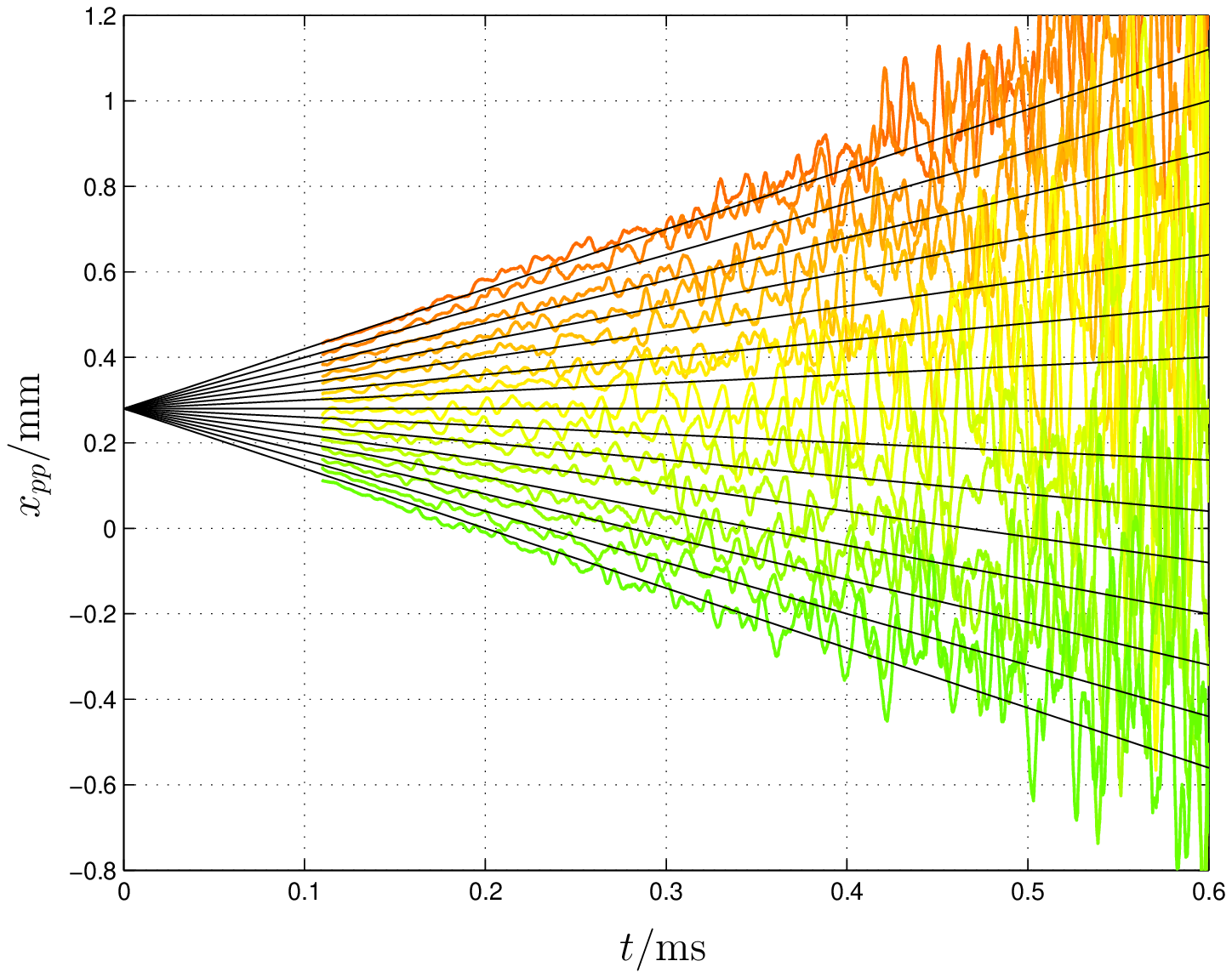}
  \caption{The same as Figure 7, but for the flow of
  GaInSn in a tube.
  }
  \label{fig:gainsn-velo}
\end{figure}

\begin{figure}
  \centering
  \includegraphics[width=8cm]{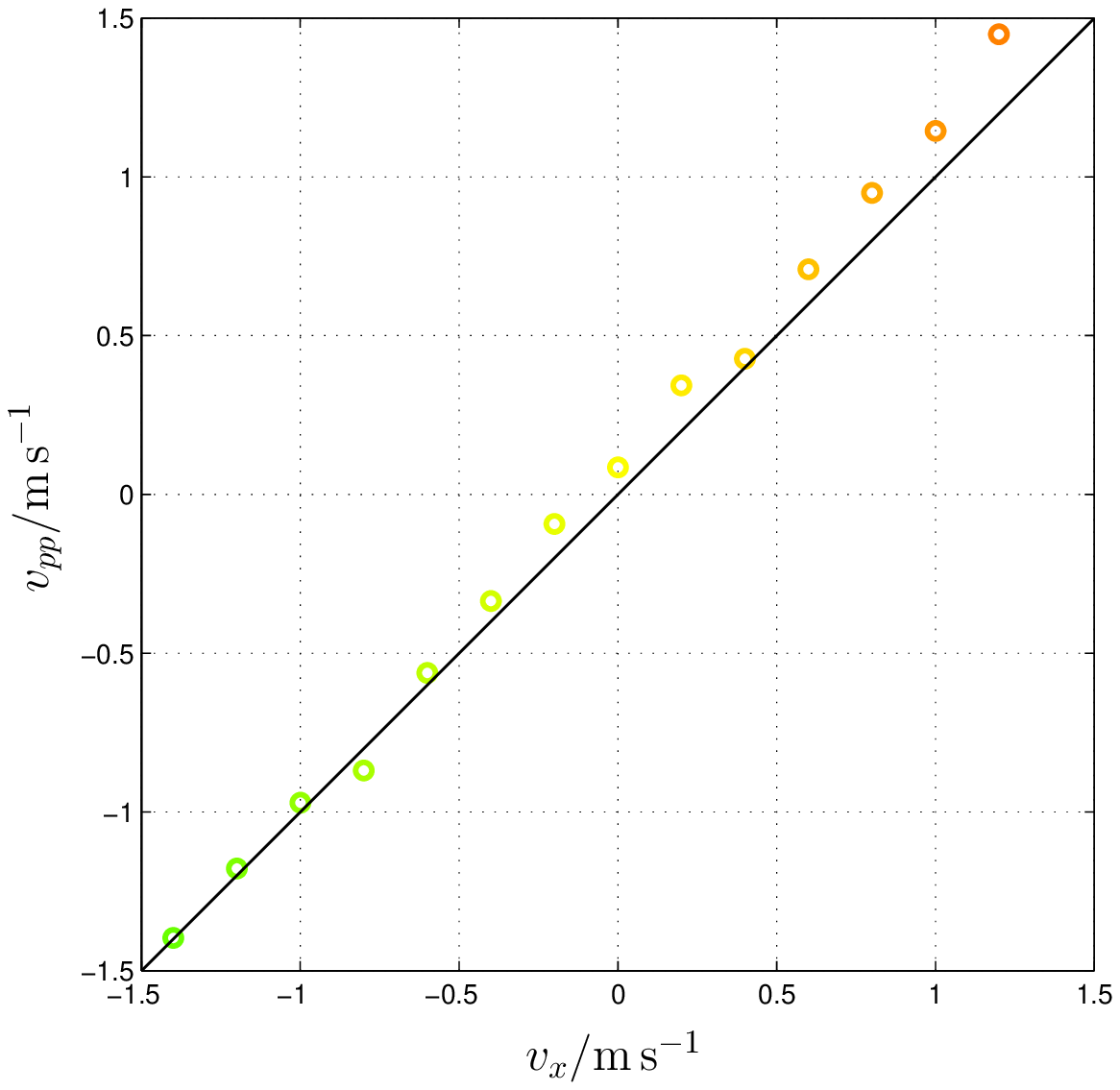}
  \caption{The same as Figure 8, but for the flow of
  GaInSn in a tube.
  }
  \label{fig:gainsn-end}
\end{figure}

Again, the time-dependence of the pole position $x_{pp}$ 
(figure 10) 
and the resulting velocities $v_pp$ are shown (figure 11).
Given the turbulent character of the pipe flow it
is not surprising that the pole trajectory in figure 10
is much more noisy than the correponding 
pole trajectory for the aluminum 
disk. Consequently, we obtain also a larger
deviation of the final velocity data (figure 11). 
Yet, the overall agreement with the pre-given velocity $v_x$ 
is quite reasonable, and might even be 
improved by a refined data analysis.

\section{Outlook}

We have modified, and significantly enhanced, 
the pulse method of Zheigur and Sermons 
\cite{ZHEIGUR} by using a compact design 
of three small detection coils embedded into 
a larger excitation coil, and by exploiting the
complete signals measured at these coils for the determination of 
the flow velocity.

This `transient eddy current flow-metering' (TEC-FM) 
has been validated at a rotating aluminum disk. First 
experiments at a flow of GaInSn show the applicability
of the method for liquid metals.

The method is calibration-free since it does not depend 
on the conductivity of the moving body (at least, as long as
the signals are still strong enough). A further advantage of the
sensor system is the 
avoidance of any magnetic material. Indeed, only air coils are 
used which can easily be adapted even to 
very hot environments such as for slab casters 
or within the pullers of Czochralski crystal growth.

A straightforward extension of the principle is to 
measure two dimensional flow structures close to the wall, 
just by adding two additional receiver coils orthogonal to 
the existing ones. 

As mentioned above, there are subtle differences between 
switching on and switching off the current. These have 
to do with the ${\bf v} \times {\bf B}$ induction term 
which adds a weak  $\infty$-shaped current system to the 
present ring-like 
one (as shown in figure 3). A detailed study of 
these differences, and of their 
potential to infer also the conductivity of the fluid, 
is left for future work.

\ack
This work was supported by Helmholtz-Gemeinschaft
Deutscher Forschungszentren (HGF) in frame  of the 
Helmholtz Alliance ''Liquid metal technologies'' 
(LIMTECH). We thank Dominique Buchenau, Hans Georg 
Krauth\"auser and J\={a}nis Priede for helpful discussions.


\section*{References}

\begin{thebibliography}{20}

\bibitem{ECKERT}  Eckert S, Buchenau D, Gerbeth G, Stefani F and Weiss FP 
2011 Some recent developments in the field of measuring techniques and instrumentation 
for liquid metal flows
{\it J. Nucl. Sci. Techn.} {\bf 48} 490-9

\bibitem{MST1}
Stefani F and Gerbeth G 2000 
A contactless method for velocity reconstruction in 
electrically conducting fluids 
{\it Meas. Sci. Techn.} {\bf 11} 758-65


\bibitem{CIFT}
Stefani F,  Gundrum T and Gerbeth G 2004 Contactless inductive flow tomography
{\it Phys. Rev. E} {\bf 70} 056306


\bibitem{PRIEDE}
Priede J,  Buchenau D and Gerbeth G 2011 
Contactless electromagnetic phase-shift flowmeter for liquid metals
{\it Meas. Sci. Techn.} {\bf 22} 055402

\bibitem{THESS}Thess A, Votyakov E V and Kolesnikov Y 2006 Lorentz force
velocimetry  {\it Phys. Rev. Lett.} {\bf 96} 164501


\bibitem{HALBEDEL}Halbedel B et al 2014 
A novel contactless flow rate measurement device for weakly 
conducting fluids based on Lorentz force velocimetry
{\it Flow Turb. Comb.} {\bf 92} 361-9

\bibitem{JULIUS}Haubrich H and Julius E 1994 
Flow meter {\it US Patent} US5426983A

\bibitem{DUBOVIKOVA}Dubovikova N, Karcher C and Kolesnikov Y 2014
Applications of Lorentz force techniques for
flow rate control in liquid metals {\it Proceedings of the 9th
PAMIR International Conference on Fundamental and Applied MHD 
(Riga - Latvia, June 16-20, 2014)} Vol 2 pp 118-122


\bibitem{ZHEIGUR}Zheigur BD and Sermons GY 1965 
Pulse method of measuring the rate of flow of a conducting fluid
{\it Magnetohydrodynamics} {\bf 1 (1)} 101-104


\bibitem{EDDYBUCH} Tegopoulos J A and Kriezis E E 1985 
{\it Eddy Currents in Linear Conducting
Media} {Elsevier: Amsterdam}

\bibitem{FORBRIGER}  Forbriger J, Galindo V, Gerbeth G and Stefani F 
2008 Measurement of the spatio-temporal 
distribution of harmonic and transient eddy currents in a liquid metal
\textit{Meas. Sci. Technol.} {\bf 19} 045704

\bibitem{BUCHENAU}
Priede J,  Buchenau D and Gerbeth G 2009 
Force-free and contactless sensor for electromagnetic
flowrate measurements
{\it Magnetohydrodynamics} {\bf 45} 451-8



\end{thebibliography}
\end{document}